\documentclass[a4paper,11pt]{article}
\usepackage{pos}

\title{Outreach, educational activities and communication of the ALICE collaboration}

\author[]{Simone Ragoni$^{a,*}$, for the ALICE collaboration}

\affiliation[a]{Creighton University, 2500 California Plaza,\\
Omaha, NE 68178, United States}

             \begin{NoHyper}
               \renewcommand\thefootnote{$\speakerFlag$}%
               \footnotetext{Speaker}%
               \stepcounter{footnote}%
               \let\thefootnote\oldthefootnote%
             \end{NoHyper}

\emailAdd{simone.ragoni@cern.ch}

\abstract{Outreach and communication with the public is an integral part of our work as researchers. A wide range of activities and platforms allow ALICE members to share, especially with the young generation, the excitement of our field. ALICE Masterclasses for high-school students, both in-person and online, are expanding, reaching a higher number of students every year. Visits to the experiment site, especially to the underground installations when the LHC schedule allows, are very popular; the large demand also serves to motivate ALICE members to get involved as guides. The surface exhibition offers a glimpse to both the physics and the variety of detectors of ALICE. Virtual visits are also popular, and the growing use of social media platforms like Instagram brings the excitement of the physics of the quark-gluon plasma to new audiences of different ages and interests.}

\FullConference{%
  42nd International Conference on High Energy Physics (ICHEP2024)\\
  18-24 July 2024\\
  Prague, Czech Republic
}

\begin{document}
\maketitle

\section{Pop-ularising particle physics}
Particle physics as a whole has often been deemed obscure and hard to understand, limited to just the small circle of the community of high energy physicists. This, in turn, ended up giving origin to a bridge between the particle physics community and the general public.  

ALICE (A Large Ion Collider Experiment) is the experiment designed to study heavy ion collisions at the CERN Large Hadron Collider (LHC). The ALICE Collaboration has developed a comprehensive outreach, educational, and communication program to demystify particle physics and engage diverse audiences worldwide. ALICE has made significant efforts to bridge the gap between its scientific endeavors and the general public, in hopes of addressing the perceived inaccessibility of particle physics.

This contribution outlines the strategies and key initiatives carried out by the ALICE collaboration to promote its research, its achievement in terms of scientific output, while also inspiring future generations of scientists, and explaining to the general public the importance and relevance of particle physics, also in everyday life. The various outreach programs, educational activities, and communication strategies used by ALICE to engage with different audiences, broadly encompassing high school and university students, fellow researchers, and the public from different age ranges, will be described.

In addition, this contribution will explain the challenges encountered in effectively communicating complex particle physics concepts to audiences without a scientific background, and the unconventional solutions developed by the ALICE collaboration to overcome these obstacles. 

\section{The importance of social media}

\begin{figure}[h]
    \centering
    \includegraphics[width=0.7\textwidth]{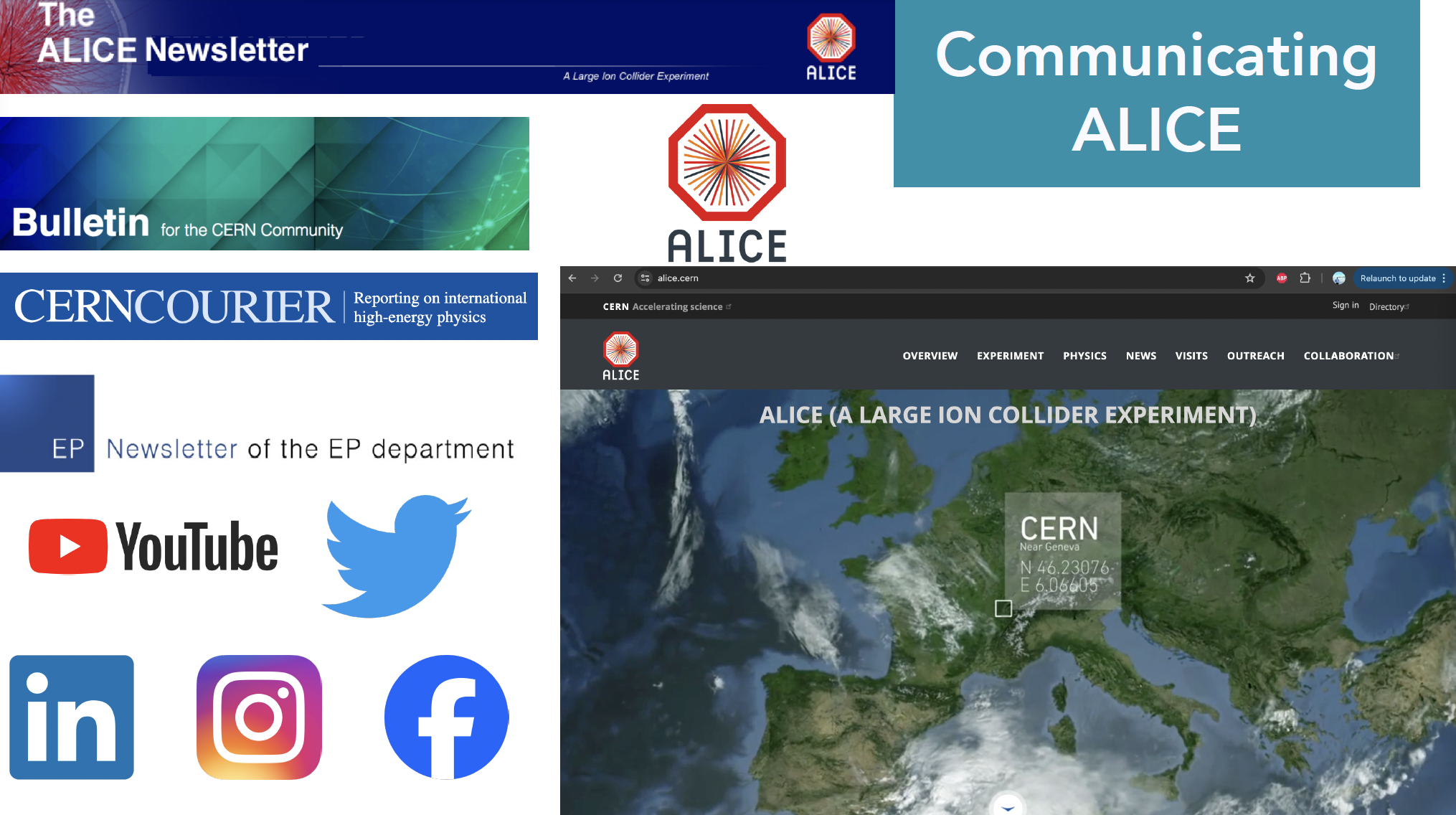}
    \caption{Communicating ALICE involves using a multitude of social media channels, each addressing different audiences.}
    \label{fig:social}
\end{figure}

The ALICE collaboration employs a strategic approach to communication, using a variety of online platforms to reach different audiences, including the general public, students and fellow physicists. Each channel is tailored to serve specific purposes and target audiences, ensuring effective reach and engagement across diverse groups, age ranges, and diverse backgrounds. The social media effort can be inspected in Fig.~\ref{fig:social}, and it is explained in detail below.

On Twitter, ALICE operates two accounts: \href{https://twitter.com/ALICEexperiment}{\texttt{@ALICEexperiment}} (handle: \texttt{@ALICEexperiment}) is intended for the general public, sharing accessible updates, news, and engaging content that highlights the work and achievements of the collaboration. The second account, \href{https://twitter.com/ALICE_Publications}{\texttt{@ALICE\_Publications}} (handle: \texttt{@ALICE\_Publications}), is targeted at fellow physicists and research institutions, focusing on sharing detailed scientific publications, technical results, and discussions relevant to the academic and research community.

The new \href{https://alice.cern/}{ALICE website} serves as the central hub of information, accessible to the general public, educators, students and the scientific community as well. It provides comprehensive details about the experiment, its goals, scientific results, and educational resources. 

\href{https://www.instagram.com/alice_experiment}{Instagram} (handle: \texttt{@alice\_experiment}) and \href{https://www.facebook.com/AliceExperiment}{Facebook} (handle: \texttt{@ALICE.experiment}) are primarily used to engage the general public with visually appealing content, such as images and videos  showcasing the day-to-day work of the collaboration, behind-the-scenes activities, and significant events. These platforms are instrumental in making the complex world of particle physics more relatable and interesting to a broad audience.

On its \href{https://www.youtube.com/@ALICEexperiment}{YouTube} channel (handle: \texttt{@ALICEexperiment}) instead, ALICE shares educational videos and interviews.

The \href{https://www.linkedin.com/company/alice-cern/}{LinkedIn} (profile: \texttt{ALICE collaboration at CERN}) account is the newest addition to the social media program of ALICE, targeting at professionals, academics, and students in physics and related fields, by leveraging on the network of current and previous ALICE members, while also showcasing ALICE's scientific achievements and career opportunities.

The \href{https://alice-collaboration.web.cern.ch/newsletter}{ALICE Newsletter} is an internal communication tool, distributed exclusively to members of the ALICE collaboration on a weekly basis. It provides detailed updates on internal events, the latest results, collaboration-specific news, and the latest career opportunities, thus fostering a sense of community and ensuring that all members are informed about the latest developments.

Additionally, ALICE engages with the broader particle physics community through publications such as the \href{https://ep-news.web.cern.ch/}{CERN EP Newsletter}, the \href{https://www.home.cern/resources/bulletin/cern/cern-bulletin}{CERN Bulletin}, and the \href{https://cerncourier.com/}{CERN Courier}. These channels are aimed at fellow physicists and researchers, offering in-depth articles, technical discussions, and the latest news from across CERN, including insights from the ALICE collaboration.

By effectively using the combination of all these channels, the ALICE collaboration is able to tailor its communication to the needs and interests of different audiences, ensuring broad engagement and spread of its scientific goals and achievements.

\section{ALICE Visit Program}

The ALICE collaboration has developed a comprehensive visit program, offering various modes of interaction, thus enabling visitors from different geographical locations and with diverse interests to experience ALICE in a manner that best suits them. The visit program is divided into three modes: real visits, virtual visits, and Instagram LIVE sessions.

Real visits to ALICE provide a tangible and immersive experience for visitors. These visits are typically conducted in three distinct areas:

\begin{enumerate}
    \item \textbf{ALICE Control Room:} Visitors begin their journey at the ALICE Control Room, where they can observe the operations team in action. Here, they gain insight into the day-to-day activities involved in monitoring and controlling the ALICE detector, along with the technical challenges in keeping the data taking volumes, and here they can really appreciate the upgrades from Run 2 to Run 3 of ALICE;
    
    \item \textbf{ALICE Exhibition Centre:} After the Control Room, visitors proceed to the ALICE Exhibition Centre, shown in Fig.~\ref{fig:exhib}, where they are introduced to the experiment’s scientific goals and the underlying principles of particle physics. The exhibition features informative displays, detector exhibits, and multimedia presentations that cater to both novices and those with a background in science;
    
    \begin{figure}[h]
    \centering
    \includegraphics[width=0.7\textwidth]{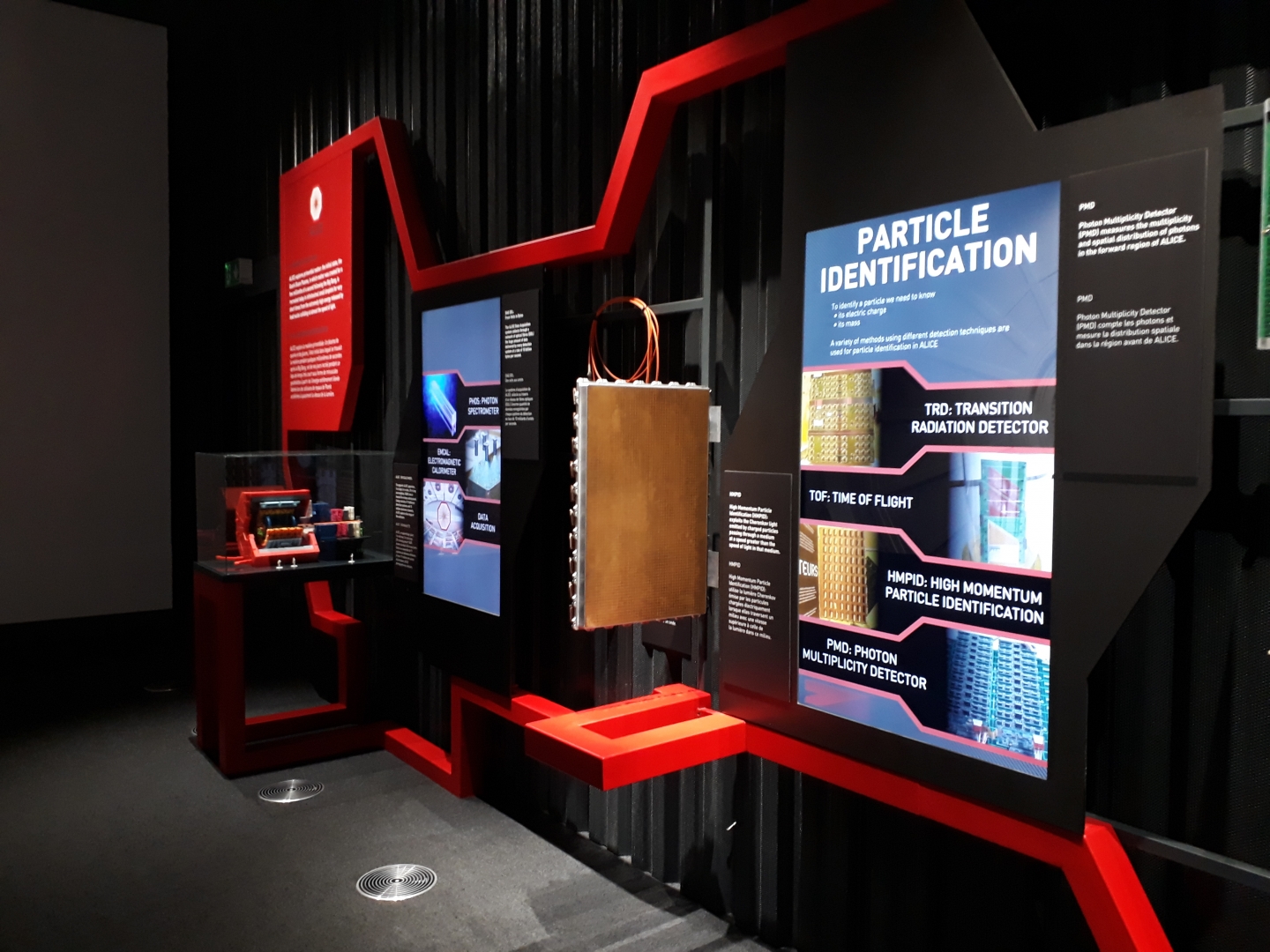}
    \caption{The new ALICE Exhibition centre.}
    \label{fig:exhib}
\end{figure}

    \item \textbf{ALICE Cavern:} The visit culminates with a descent into the ALICE Cavern, where the massive detector is housed. This portion of the visit allows guests to appreciate the scale and engineering marvel of the ALICE experiment up close. Guided tours provide detailed explanations of the detector components and their roles in the data collection process, offering a unique perspective on the infrastructure that supports cutting-edge research in particle physics.
\end{enumerate}

Recognizing the need to make ALICE accessible to a global audience without geographical constraints, the collaboration also offers virtual visits via Zoom, which became more popular after 2020 during the COVID pandemic. Virtual visits typically include a live tour of the ALICE Exhibition Centre, virtual access to the Control Room, and a guided walkthrough of the ALICE Cavern. These visits are hosted by a member of the ALICE team, aiming to provide an engaging and informative experience that mirrors the structure of the real visits as closely as possible, followed by a Q\&A session.

The latest addition to the ALICE visit program is the Instagram LIVE series, which has quickly become a popular platform for engaging with a broader and more diverse audience, with just the aid of your own smartphone. These sessions are conducted by the ALICE Social Media team, who host live discussions on various topics related to the experiment and particle physics in general. Each session is designed to be accessible to all, with a strong focus on promoting diversity and inclusivity. The Instagram LIVE format allows for a more casual and interactive atmosphere, where viewers can ask questions and participate in discussions in real-time. Topics are carefully selected to appeal to a wide audience, from basic introductions to particle physics to more in-depth explorations of specific research findings or technical aspects of the ALICE experiment, usually gathering about 500 to 600 people per hour of LIVE session.

The ALICE visit program, with its combination of real visits, virtual tours, and Instagram LIVE sessions, ensures that the wonder and complexity of particle physics are accessible to everyone, regardless of their location or background. By providing multiple modes of engagement, ALICE is able to reach a diverse audience, inspiring curiosity and fostering a deeper understanding of the scientific work being conducted at CERN.

\section{ALICE Masterclass Program}

The ALICE Masterclass program is a cornerstone of the collaboration's educational outreach, offering students a unique opportunity to engage with real data from the ALICE experiment. It is part of the "International Masterclasses in Particle Physics" program \cite{PhysicsMasterclasses},  the flagship activity of the International Particle Physics Outreach Group (IPPOG) \cite{IPPOG_Masterclasses} ALICE is a member of. This program is designed to provide hands-on experience in particle physics, enabling students to perform data analysis.

\subsection{The original ALICE Masterclass}

The original incarnation of the ALICE Masterclass requires participants to install the ROOT analysis package \cite{ROOT}. This version of the Masterclass offers two distinct activities:

\begin{enumerate}
    \item \textbf{Measurement of \(R_{AA}\):} This exercise allows students to explore the nuclear modification factor \(R_{AA}\), a critical observable in the study of quark-gluon plasma and the behavior of matter under extreme conditions. Participants analyse collision data to understand how particle production is suppressed in heavy-ion collisions compared to proton-proton collisions.
    
    \item \textbf{Strangeness Enhancement:} In this activity, students investigate the phenomenon of strangeness enhancement, a key indicator of the formation of quark-gluon plasma. By analysing the production of strange particles in heavy-ion collisions, students gain insights into the unique properties of matter in the early universe.
\end{enumerate}

While the traditional Masterclass provides a deep and immersive experience, its requirement for ROOT installation poses a significant barrier, limiting accessibility, particularly for schools and institutions without the specific technical support.

\subsection{The new Web-Based ALICE Masterclass}

To overcome these challenges, the ALICE collaboration has introduced a new, web-based version of the Masterclass, accessible through the \href{https://alice-web-masterclass.app.cern.ch/home}{ALICE Masterclass website}. This updated platform offers the strangeness enhancement exercise, with a vastly improved accessibility and a more immediate experience with the respect to the original Masterclass.

The primary advantage of the web-based Masterclass is its simplicity and ease of use. Participants only need access to a web browser, eliminating the need for any software installation. Hence, it significantly lowers entry barriers, making it accessible to a much broader audience, including students and educators in remote or under-resourced areas. The web-based platform also facilitates participation in larger groups and makes it easier to integrate the Masterclass into school curricula or public outreach events.

The introduction of the web-based ALICE Masterclass represents a significant step forward in the collaboration's efforts to democratize access to cutting-edge scientific education. ALICE is then able to reach a global audience, inspiring the next generation of physicists and broadening public understanding of particle physics.

\section{Conclusions}

The ALICE collaboration has an outstanding commitment to outreach, education, and public engagement through a diverse set of programs. By maintaining an active online presence, ALICE ensures that its groundbreaking research in particle physics is communicated clearly and accessibly, inspiring curiosity and fostering a deeper understanding of the scientific process.

Complementing the social media efforts, the ALICE visit program provides both physical and virtual opportunities for the public, students, and educators to experience the experiment firsthand. ALICE makes it possible for people from around the world to engage with the experiment in a meaningful way, inspiring a deeper appreciation for the complexity and importance of particle physics research.

Finally, the ALICE Masterclass program stands as a flagship educational activity, particularly within the framework of the International Particle Physics Outreach Group (IPPOG) \cite{IPPOG_Masterclasses}. It provides students with the opportunity to analyse real ALICE data. By engaging directly with the scientific process, participants gain valuable insights into the world of particle physics, fostering the next generation of scientists.

In summary, ALICE’s outreach and educational initiatives are a testament to the collaboration’s dedication to making particle physics accessible and engaging to all. Through a combination of social media strategies, comprehensive visit programs, and  educational activities like the Masterclass, ALICE continues to play a pivotal role in promoting science education and public engagement on a global scale.

\end{document}